\documentclass{PoS}

\usepackage{amssymb}
\usepackage{amsmath}
\pdfoutput=1

\title{Inflation and dark energy from $f(R)$ gravity}

\ShortTitle{Inflation and DE from $f(R)$ gravity}

\author{\speaker{Micha\l{} Artymowski}\\
        Institute of Physics, Faculty of Physics, Astronomy and Applied Computer Science, Jagiellonian University ul. {\L}ojasiewicza 11, 30-348 Krak{\'o}w, Poland\\
        E-mail: \email{Michal.Artymowski@uj.edu.pl}}
        
        \author{Marek Lewicki\\
        Institute of Theoretical Physics, Faculty of Physics, University of Warsaw ul. Pasteura 5, 02-093 Warsaw, Poland\\
        E-mail: \email{Marek.Lewicki@fuw.edu.pl}}

\author{Zygmunt Lalak\\
        Institute of Theoretical Physics, Faculty of Physics, University of Warsaw ul. Pasteura 5, 02-093 Warsaw, Poland\\ 
        E-mail: \email{Zygmunt.Lalak@fuw.edu.pl}}

\abstract{Inflationary paradigm has several issues, such as the pre-inflationary horizon problem or the eternal inflation. To avoid that the standard Starobinsky inflation has been extended to $R + \alpha R^n$ and $R + \alpha R^n -\delta R^{2-n}$
models as well as Brans-Dicke generalisation of those $f(R)$ models. The region of the parameter space, which provides consistency with PLANCK data and lack of eternal inflation has been founded. The Einstein frame potential has a stable minimum with non-zero vacuum energy, which may be a source of dark energy.
}

\FullConference{Proceedings of the PLANCK 2015 conference,\\
		25-29 May 2015\\
		Ioannina, Greece }

\begin{document}

\section{Introduction}

Recent data from the PLANCK satellite \cite{Ade:2013zuv} has set new constrains on cosmic inflation \cite{Lyth:1998xn,Liddle:2000dt}. In particular new constrains on the tensor-to-scalar ratio $r$ seems to prefer plateau potentials of e.g. Higgs \cite{Bezrukov:2010jz} or Starobinsky \cite{Starobinsky:1980te} inflation, over the power-law potentials of chaotic inflationary models, like $m^2\phi^2$. Potentials with inflationary plateau are on one hand side perfectly consistent with observational data, but on the other hand they suffer from a series of theoretical issues \cite{Ijjas:2013vea}, for instance from problems of initial conditions and eternal inflation. 

The first problem originates from the fact that the inflationary plateau is limited from above by the GUT scale, which means that at the Planck scale the contribution of the inflatons potential energy density to the total energy density of any pre-inflationary horizon is negligible. Because of that inhomogeneities of the inflaton (which may have significant contribution to the total energy density at the Planck era) may dominate the system before inflation will have a chance to appear, which may spoil the inflationary paradigm. To avoid this problem, one needs to assume that the inflaton's energy density is quite homogeneous inside of the region of order of billions of horizons at the Planck scale, which is the source of the pre-inflationary horizon problem

The other issue is, that potentials with plateau support existence eternal inflation. For sufficiently large values of inflaton $\phi$ one obtains a situation in which quantum fluctuations dominate the evolution of $\phi$. During each Hubble time from one horizon inflation generates $\sim20$ horizons, which keep inflating. Thus, for quantum fluctuations domination there are always horizons in which the field is not going to reach its minimum and therefore the graceful exit will not occur. If one would consider volume as a measure of probability then it would be highly unlikely to find an observer inside a horizon in which accelerated expansion has stopped. 

To avoid those problems we have presented inflation based on the Brans-Dicke model motivated by the $R + \alpha R^n$ Lagrangian density (with $n\lesssim 2$)  \cite{Artymowski:2014gea,Artymowski:2014nva}, which is a small deviation from the Starobinsky $R + \alpha R^2$ model. We have also extended our model into $R + \alpha R^n - \beta R^{2-n}$, which provides dark energy (DE) and stable vacuum for the inflaton.

\section{The $R + \alpha R^n$ model}

Let us assume that the universe may be described by the flat FRW metric and $f(R)$ Lagrangian density. Any $f(R)$ theory is equivalent to the Brans-Dicke model with auxiliary field $\varphi:=F(R)$ (where $F:=f_R:=df/dR$) non-minimally coupled to gravity, potential $U(\varphi):=(R\,F-f)/2$ and Brans-Dicke parameter $\omega = 0$. In order to seek for a part of parameter space which solves problems of inflationary plateau potentials let us generalise the $f(R) = R + \alpha R^n$ into Brans-Dicke theory with any value of $\omega$. In such a case the Jordan frame equations of motion take the following form
\begin{eqnarray}
\ddot{\varphi} + 3H\dot{\varphi} + \frac{2}{\beta}(\varphi U_\varphi - 2U) &=& \frac{1}{\beta}\left(\rho_M - 3P_M\right)\ ,\label{eq:motionBD}\\
3\left(H + \frac{\dot{\varphi}}{2\varphi}\right)^2 &=& \frac{\beta}{4}\left(\frac{\dot{\varphi}}{\varphi}\right)^2 + \frac{U}{\varphi} + \frac{\rho_M}{\varphi}\, , \label{eq:FriedBD}\\
\dot{\rho}_M + 3H(\rho_M + P_M) &=& 0\, , \label{eq:cont}
\end{eqnarray}
where $\beta = 2\omega+3$, $U_\varphi:=\frac{dU}{d\varphi}$ and $\rho_M$ and $P_M$ are energy density and pressure of matter fields respectively. The Jordan frame scalar potential of the $R + \alpha R^n$ model takes the form 
\begin{equation}
U = \frac{n-1}{2}\alpha\left(\frac{\varphi-1}{n\alpha}\right)^\frac{n}{n-1} \, . \label{eq:Uinfl}
\end{equation}
\\*

\begin{figure}[ht]
\begin{minipage}[t]{0.47\linewidth}
\centering
\includegraphics[scale=0.66]{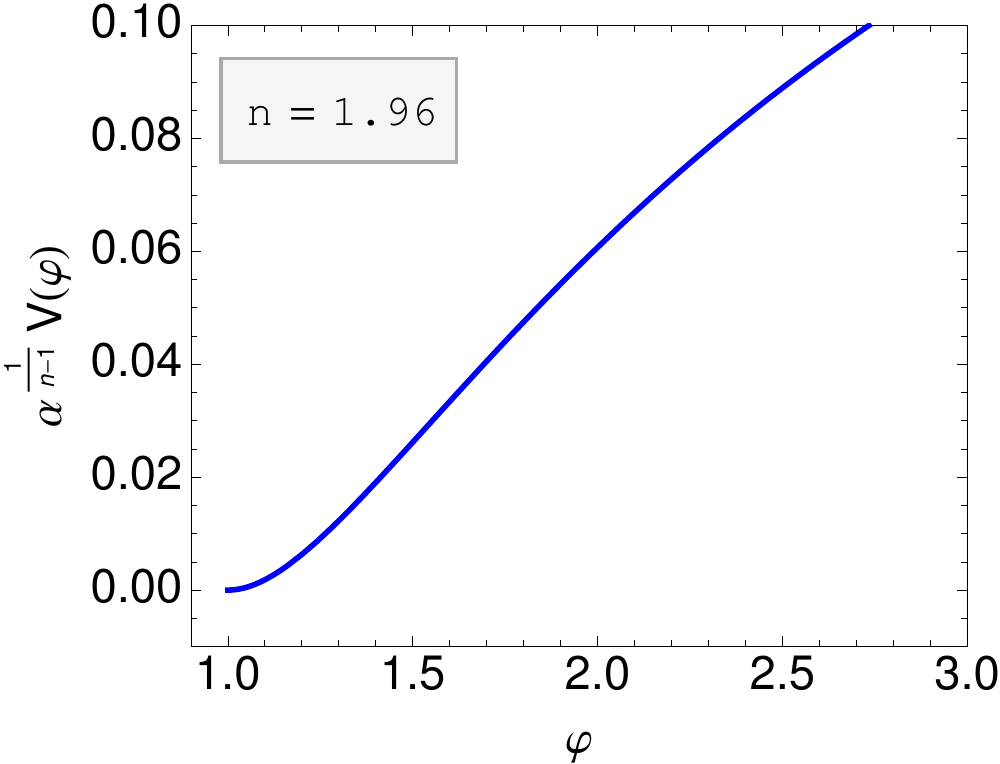}
\end{minipage} 
\begin{minipage}[t]{0.47\linewidth}
\centering
\includegraphics[scale=0.84]{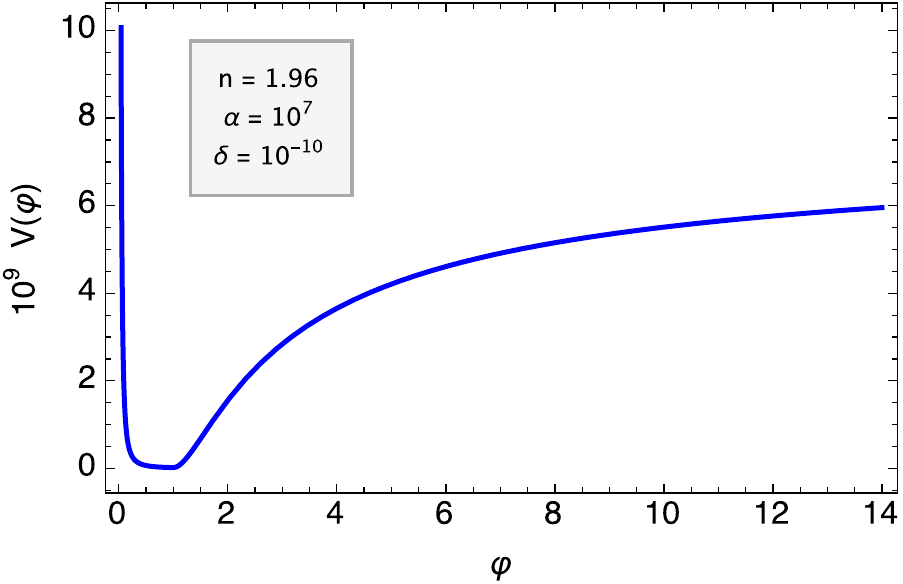}
\end{minipage} 
\caption{Left/right panels: The Einstein frame potential as a function of the Jordan frame field for models without/with dark energy. Inflationary parts of both potentials are the same, but the presence of dark energy contribution stabilise the GR vacuum.}
\label{potentials}
\end{figure}

\begin{figure}[ht]
\begin{minipage}[t]{0.47\linewidth}
\centering
\includegraphics[scale=0.68]{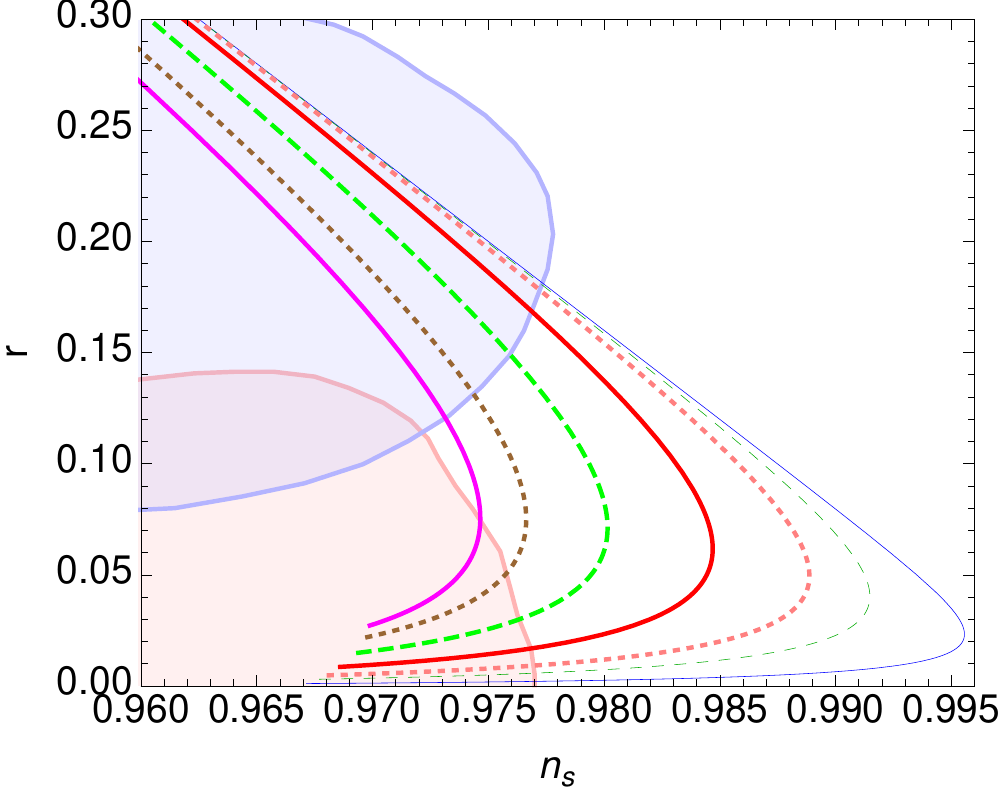}
\end{minipage} 
\begin{minipage}[t]{0.47\linewidth}
\centering
\includegraphics[scale=0.68]{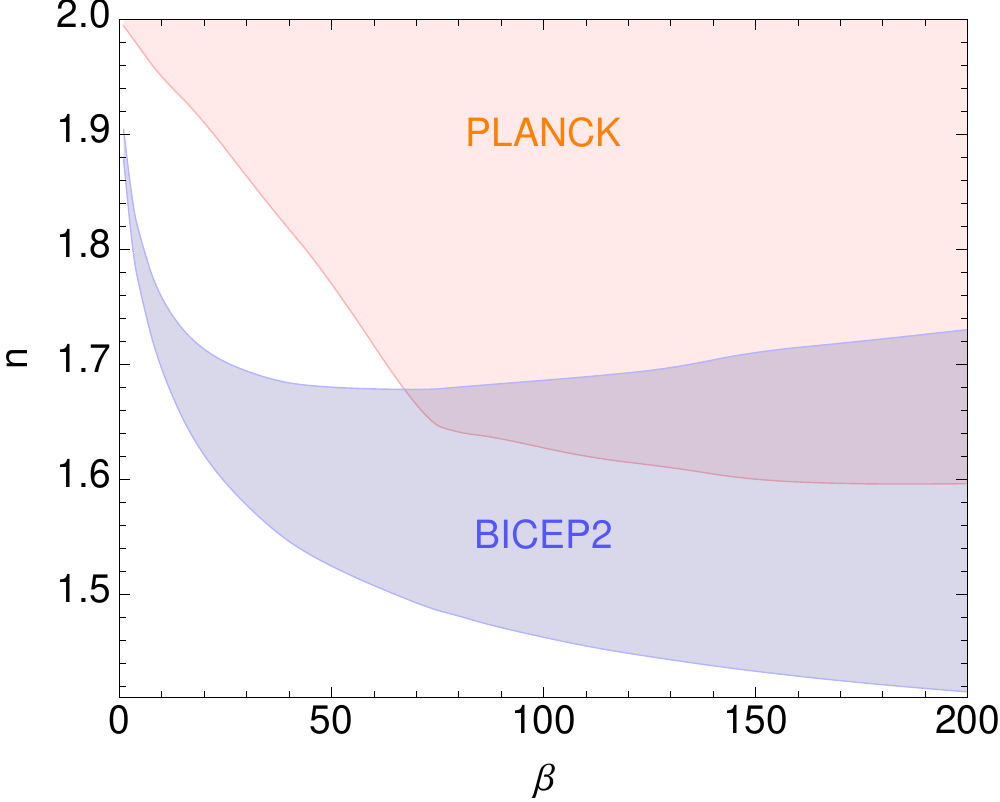}
\end{minipage} 
\caption{Left panel: The $(r,n_s)$ plane for different $n$ and for $\beta = 1$ (blue solid line), $\beta = 3$ (green dashed line line), $\beta=5$ (dotted orange line), $\beta = 10$ (red line), $\beta = 20$ (light green dashed line), $\beta = 35$ (brown dotted line) and $\beta = 50$ (pink line). Blue and Red regions represent $2\sigma$ regimes of BICEP2 and PLANCK data respectively. Right panel: Regions on the $(\beta,n)$ plane for which the primordial inhomogeneities obtained for the generalisation of the $R + \alpha R^n$ fit the Planck (red region) or BICEP2 (blue region) data.
}
\label{data}
\end{figure}

The gravitational part of the action may obtain its canonical (minimally coupled to $\varphi$) form after transformation to the Einstein frame. Then for the Einstein frame metric tensor
\begin{equation}
\tilde{g}_{\mu\nu}=\varphi g_{\mu\nu}\, , \qquad d\tilde{t}=\sqrt{\varphi}dt\, ,\qquad\tilde{a} = \sqrt{\varphi}a
\end{equation}
one obtains the action of the form of
\begin{equation}
S = \int d^4x \sqrt{-\tilde{g}}\left[ \frac{1}{2}\tilde{R} - \frac{1}{2}\left( \tilde{\nabla}\phi \right)^2 - V(\phi)\right]\, ,
\end{equation}
where $\tilde{\nabla}$ is the derivative with respect to the Einstein frame coordinates. The Einstein frame scalar field $\phi$ and potential $V$ are defined by
\begin{equation}
\phi = \sqrt{\frac{\beta}{2}}\log\varphi \, , \qquad \varphi = \exp\left(\sqrt{\frac{2}{\beta}}\phi\right) \,  , \qquad V = \left.\frac{U(\varphi)}{\varphi^2}\right|_{\varphi=\varphi(\phi)} \, .
\end{equation}
The Einstein frame potential of the (\ref{eq:Uinfl}) model has been presented in the left panel of Fig. \ref{potentials}. One can see that for big $\varphi$ the potential is quite flat and capable of generating inflation, but the potential has no minimum, which would be responsible for the graceful exit and the reheating of the universe. Nevertheless Just from the shape of the potential one can see two advantages of this model over the Starobinsky inflation. First of all $V$ is not limited from above, which means that the potential may dominate energy density inside some of pre-inflationary horizons. Therefore taking $n\lesssim 2$ solves the pre-inflationary horizon problem. Second of all one can require quantum fluctuations of the inflaton to be small comparing to classical evolution of $\phi$ from the scale to inflation up to the Planck scale. As shown in Ref. \cite{Artymowski:2014nva} this gives the following constrain on $\beta$
\begin{equation}
\beta < 24\pi^2\left(\frac{n-2}{n-1}\right)^2 \, . \label{eq:BetaCondition}
\end{equation}
This means that the problem of eternal inflation may be solved in this model in certain part of the parameter space. Again, the help comes from the fact that the inflationary potential is not as flat as in the Starobinsky model, which suppresses quantum fluctuations effects. Fig. \ref{data} shows predictions of the model at the $(r,n_s)$ plane as well as the allowed range of the parameter space, which fits to the data.

\section{The $R + \alpha R^n - \beta R^{2-n}$ model}

As mentioned in the previous section the $R + \alpha R^n$ model suffers from the lack of minimum of the Einstein frame potential. To solve that problem let us introduce an additional term to the $f(R)$ function, namely
\begin{equation}
f(R) = R + \alpha R^{n} - \delta R^{2-n} \, ,
\end{equation}
where $\alpha, \beta>0$ and $\beta,\alpha\, \beta \ll 1$. In such a case the potential obtains the form from the right panel of Fig. \ref{potentials}. The minimum is now obtained for all values of $n$ and $\beta$, if one would decide to extend this model into a Brans-Dicke theory. Then one obtains
\begin{equation}
U(\varphi) = \frac{1}{2} (n-1)\left(\alpha  \mathbb{R}^{ n}(\varphi )+\delta\mathbb{R}^{2-n}(\varphi )\right) \, ,\quad  V = \frac{1}{\varphi^2}U \, ,\label{eq:DEpotential}
\end{equation}
where
\begin{equation}
\mathbb{R}(\varphi) = \left(\frac{\sqrt{4 (2-n)n \alpha  \delta +(\varphi -1)^2}+\varphi -1}{2n \alpha }\right)^{\frac{1}{n-1}} \, .
\end{equation}
The $\mathbb{R}$ function could be interpreted as the Ricci scalar in the $f(R)$ theory. However in the considered model (for $\beta\neq 3$) it has no connection with the curvature. The Einstein frame potential has a minimum at  $\varphi_{min} \simeq \frac{2}{n} (n-1) (1+2 n \alpha  \delta )$. The value of $V$ at the minimum for small values of $\delta$ reads
\begin{equation}
V(\varphi_{min})\simeq \frac{n}{8 (n-1)^2} (n \delta )^{\frac{1}{n-1}} \left(n-1-n^2 \alpha  \delta \right) \sim \frac{1}{2}\delta^{\frac{1}{n-1}} \, .\label{eq:Vmin}
\end{equation}
Hence, this model predicts some amount of vacuum energy. If one would express $\delta$ in term of a power of some mass scale $M$ then the vacuum energy would be of order of $M^{2}$ (in Planck units). In order to fit the DE data one requires $M\sim 10^{-60} M_p$. This mass scale is significantly below the physics of the Standard Model, but this troublesome gap is typical for most of dark energy models. 

\begin{figure}[h]
\centering
\includegraphics[height=4.6cm,bb=0 0 260 167]{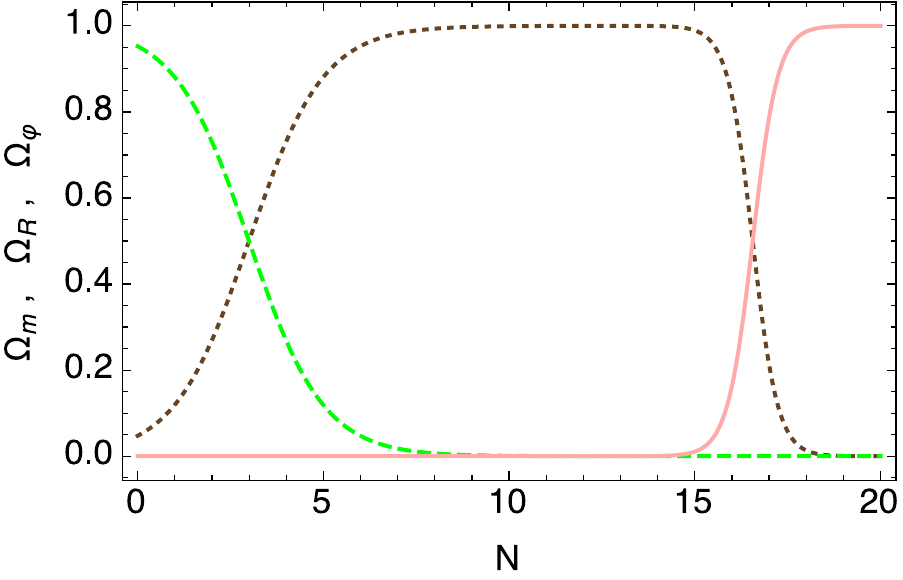}
\caption{\it Numerical result of the evolution of {density parameters $\Omega_m$ (dust), $
\Omega_R$ (radiation) and $\Omega_\varphi$ (scalar field, dotted brown, dashed green and pink lines respectively) as a function of N for $n=1.95$, $\alpha = 2\times 10^8$, $\delta = 10^{-30}$ and $\beta = 10$}. Note that the vacuum energy of the Brans-Dicke field starts to dominate when the field reaches the minimum.}
\label{fig:HDE}
\end{figure}

\section{Summary}

We have presented several issues related to Starobinsky and Higgs models of cosmic inflation, namely: 1) the pre-inflationary horizon problem, which originates from the fact that the Einstein frame potentials of those models are limited from above by the GUT scale, 2) the problem of eternal inflation, which leads to the conclusion that in plateau models there are still many horizons which never ended inflating and therefore it is very unlikely to be in our part of the universe. 

Both of those problems may be solved by the Brans-Dicke model based on the $f(R) = R + \alpha R^n$ theory. In such a case the Einstein frame potential is exponentially growing in the big fields limit and therefore it is not limited from above. Second of all, for certain regions of the $(n,\beta)$ parameter space one obtains lack of quantum fluctuations domination up to the Planck scale. This protects us from the eternal inflation problem

The $f(R) = R + \alpha R^n$ model suffers from lack of minimum of the Einstein frame potential, which may be solved by adding an additional term to the Lagrangian density. Any Brans-Dicke model based on $R + \alpha R^n - \beta R^{2-n}$ has a stable minimum and inflationary part, which is exactly the same as in the  $f(R) = R + \alpha R^n$ scenario. The only difference is that the potential has non-zero vacuum energy density, which may play a role of dark energy, responsible for current acceleration of the universe. 

\begin{center}
{\bf Acknowledgements}
\end{center}
This work was supported by the Foundation for
European Regional Development Fund.\\This work was partially supported by National Science Centre, Poland 
under \\ research grants DEC-2012/04/A/ST2/00099 and DEC-2014/13/N/ST2/02712 and  grant FUGA UMO-2014/12/S/ST2/00243.\\
ML was supported by the Polish
National Science Centre under doctoral scholarship number
2015/16/T/ST2/00527

\end{document}